\documentclass[sigconf]{acmart}
\AtBeginDocument{%
  }

\setcopyright{acmlicensed}
\copyrightyear{2018}
\acmYear{2018}
\acmDOI{XXXXXXX.XXXXXXX}
\acmConference[Conference acronym 'XX]{Make sure to enter the correct
  conference title from your rights confirmation emai}{June 03--05,
  2018}{Woodstock, NY}
\acmISBN{978-1-4503-XXXX-X/2018/06}

\begin{document}

\title{Towards Sustainable Growth: A Multi-Value-Aware Retrieval Framework for E-Commerce Search}

\author{Yifan Wang}
\authornote{Both authors contributed equally to this paper.}
\affiliation{%
  \institution{Taobao \& Tmall Group of Alibaba}
  \city{HangZhou}
  \country{China}
}
\email{wangyifan.ivan@alibaba-inc.com}

\author{Yixuan Wang}
\authornotemark[1]
\affiliation{%
  \institution{Taobao \& Tmall Group of Alibaba}
  \city{HangZhou}
  \country{China}}
\email{suqin.wyx@alibaba-inc.com}

\author{Yidan Liang}
\authornote{Corresponding author.}
\affiliation{%
  \institution{Taobao \& Tmall Group of Alibaba}
  \city{HangZhou}
  \country{China}
}
\email{liangyidan.lyd@alibaba-inc.com}

\author{Qiang Liu}
\affiliation{%
  \institution{Taobao \& Tmall Group of Alibaba}
  \city{HangZhou}
  \country{China}
}
\email{liuqiang.lq@alibaba-inc.com}

\author{Fei Xiao}
\affiliation{%
  \institution{Taobao \& Tmall Group of Alibaba}
  \city{HangZhou}
  \country{China}
}
\email{guren.xf@alibaba-inc.com}


\renewcommand{\shortauthors}{WANG et al.}

\begin{abstract}
    New item growth is critical for maintaining a healthy ecosystem in large-scale e-commerce platforms. However, existing systems tend to prioritize presenting users with already popular items---a phenomenon often referred to as the ``Matthew effect''. In the context of search retrieval, current cold-start models suffer from the misalignment between training objectives and online business metrics, and they lack effective mechanisms to measure an item’s growth potential. In this paper, we propose a Multi-Value-Aware retrieval framework tailored for e-commerce search, designed to better align with the cascaded online values across different stages of the search system while balancing immediate conversion and long-term item growth. Our framework \textbf{GrowthGR} consists of two key components: an \textit{Item Long-term Transaction Value Prediction (ItemLTV)} module and a \textit{Multi-Value-Aware Generative Retrieval (MultiGR)} module. First, in the ItemLTV module, we employ counterfactual inference to quantify the long-term value increment attributable to a single user interaction. Second, in the MultiGR module, building upon a semantic-ID-based generative retrieval architecture, we leverage structured samples with the search cascade signals and adopt a \textit{Multi-Value-Aware Policy Optimization (MoPO)} training paradigm to align with multi-stage online values, while explicitly balancing short-term transactional value and long-term growth potential estimated by ItemLTV. We successfully deployed GrowthGR on Taobao’s production platform, achieving a substantial 5.3\% lift in new item GMV while delivering a non-trivial 0.3\% gain in overall search GMV. Extensive online analysis and A/B testing demonstrate its positive impact on the overall ecosystem value.
\end{abstract}





\maketitle


\section{Introduction}

In the era of rapidly evolving e-commerce, large-scale platforms like Taobao serve not only as transaction hubs but also as dynamic ecosystems where the continuous influx of new items is vital for long-term sustainability~\cite{wang2018billion,zhang2025cold,shen2025aliboost}. New items are the lifeblood of sustainable growth; they represent the latest consumer trends, foster brand innovation, and prevent the platform's inventory from stagnating. However, effectively distributing these new items within the search scenario remains a non-trivial challenge. Most industrial search engines are inherently optimized for historical engagement metrics, such as Click-Through Rate (CTR) and Conversion Rate (CVR)~\cite{ma2018entire}. This optimization logic tends to induce the ``Matthew Effect'' (or the ``rich-get-richer'' phenomenon), where the search system disproportionately favors head items with abundant historical data and high exposure~\cite{fabbri2022exposure}. Consequently, new items characterized by their massive scale and extreme data sparsity often struggle to break through this feedback loop~\cite{wang2018billion}. The lack of prior interactions makes it difficult for traditional retrieval models to accurately estimate their potential value, resulting in a "cold-start" dilemma that undermines the sellers' motivation and hinders the discovery experience for users. 

Existing works have attempted to tackle the cold-start challenge from two primary directions. The first focuses on representation enhancement, leveraging multi-modal information, knowledge graphs, or Large Language Models (LLMs) to inject prior knowledge and mitigate popularity bias~\cite{elkahky2015multi,he2023large,wu2024could}. The second direction involves heuristic exposure mechanisms, such as dedicated traffic quotas or manual boosting interventions, to accelerate the growth of new items. While these methods provide temporary relief, they often suffer from significant efficiency trade-offs and lack a fundamental understanding of growth. We identify two critical limitations in these prior approaches:
\begin{enumerate}
    \item \textbf{Lack of quantitative measurement for the long-term value of new items.} Most systems equate growth with simplistic metrics like current exposure share or immediate CTR/CVR, failing to capture the ripple effect an early-stage interaction has on an item's future lifecycle.
    \item  \textbf{Limitations in balancing immediate conversion efficiency with long-term ecosystem health.} 
    Existing strategies often focus too heavily on immediate gains, favoring high-conversion popular items to protect current revenue. This narrow focus sacrifices the growth of new items.
\end{enumerate}

\begin{figure}[t]
  \centering
  \includegraphics[width=\linewidth]{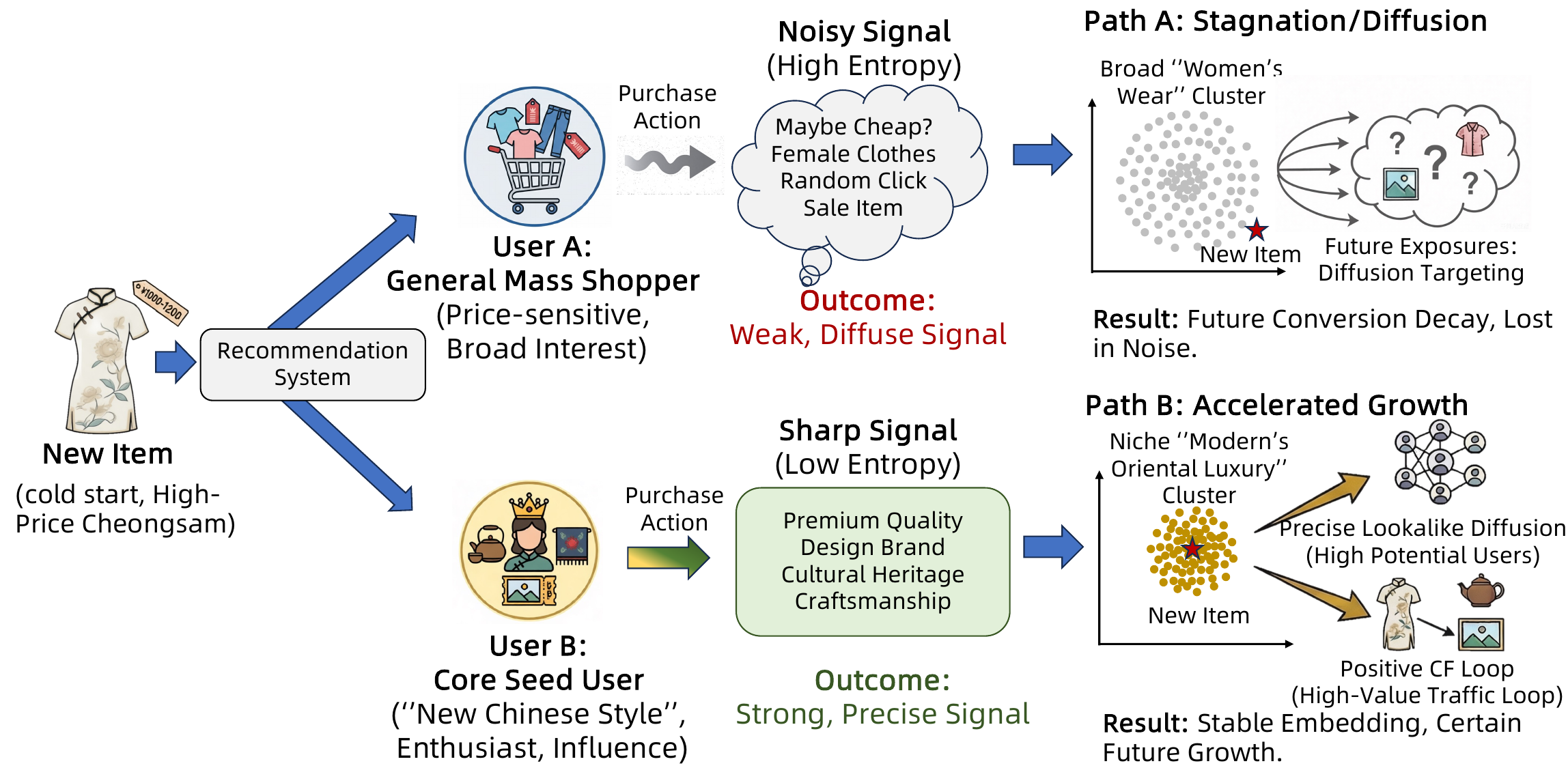}
  \caption{The Cold-start Dilemma: Immediate Conversion vs. Long-term Item Growth Value.}
  \label{intro_matthew}
\end{figure}

Inspired by these limitations, we posit that an effective item cold-start mechanism necessitates two fundamental capabilities:
\begin{enumerate}
    \item \textbf{Quantifying the marginal growth value of individual interactions}: Rather than focusing solely on immediate conversion, the system should evaluate how a single interaction contributes to the long-term transaction of a new item. This enables fine-grained traffic allocation decisions.
    \item \textbf{Precision-targeted initial distribution}: New items suffer from a lack of posterior feedback and insufficiently trained ID embeddings. Consequently, the initial trajectory is critical. Precise delivery to high-affinity users mitigates signal dilution and accelerates the acquisition of informative feedback.
\end{enumerate}
To illustrate, as shown in Figure~\ref{intro_matthew}, consider a high-end designer dress in Taobao’s cold-start scenario: The system favors a niche enthusiast over a mass consumer, even if their predicted CVRs are identical. The interaction from an enthusiast provides a significantly sharper signal, allowing the system to decode the item's precise market positioning early on. By establishing this high-fidelity initial seed, the system can leverage lookalike algorithms to accurately expand to similar high-potential users, carving out a sustainable growth path instead of being submerged in noisy, irrelevant traffic.

To address these challenges, we propose \textbf{GrowthGR}, a novel Multi-Value-Aware Generative Retrieval framework designed for sustainable new item growth. GrowthGR consists of two synergistic modules: the \textit{Item Long-term Transaction Value (ItemLTV)} module and the \textit{Multi-Value-Aware Generative Retrieval (MultiGR)} module. Recognizing that a single early-stage interaction for a new item can significantly shift its future trajectory—both in terms of system-level visibility and consumer perception—we formulate the long-term value estimation as a counterfactual causal inference problem. The ItemLTV module quantifies the uplift effect of an individual interaction on future transaction increments, providing a rigorous metric for long-term potential. Building upon this, the MultiGR module introduces a generative modeling paradigm with dense semantic priors, leveraging the semantic IDs to transcend the limitations of traditional ID-based retrieval. To bridge the gap between multiple objectives, we employ preference alignment techniques \textit{Multi-Value-Aware Policy Optimization (MoPO)} that integrate the item's cascaded values within the search funnel (from candidates, impressions, and clicks to final conversions) with its predicted long-term potential. By doing so, GrowthGR aims to make the retrieval process not only aware of immediate gains but also strategically aligned with the platform’s sustainable development.

The main contributions of this paper are summarized as follows:

\begin{itemize}

\item \textbf{A Novel Perspective on Sustainable Growth}: We shift the focus from traditional short-term efficiency to the long-term ecological value of new items. We provide a quantitative measure the ItemLTV to estimate the uplift effect of individual interactions on an item's future transaction trajectory.

\item \textbf{A Multi-Value-Aware Generative Retrieval Framework}: We propose GrowthGR, the first generative retrieval framework tailored for new item growth.

\item \textbf{Strategic Value Alignment}: We introduce a preference alignment mechanism MoPO that harmonizes multiple objectives within the search system. This allows the model to balance immediate conversion value with the predicted long-term potential.

\item \textbf{Large-Scale Industrial Validation}: We have deployed GrowthGR on Taobao, one of the world's largest e-commerce platforms. The results of large-scale online A/B tests conducted over two months indicate that our approach achieves a substantial 5.3\% lift in new item GMV while delivering a non-trivial 0.3\% gain in overall search GMV.

\end{itemize}

\section{Related Works}

\subsection{Cold-Start Recommendation}
To mitigate the data sparsity issue in cold-start scenarios, recent literature has shifted from heuristic approaches to deep learning paradigms that exploit auxiliary knowledge transfer, graph structures, and generative capabilities~\cite{chen2022generative,jiang2023self,liu2023uncertainty}. In the domain of meta-learning, 
FORM~\cite{sun2021form} innovates with an online regularized meta-leader to adapt to streaming user preferences, while PNMTA~\cite{pang2022pnmta} and PAM~\cite{luo2025online} refine this by dynamically modulating network parameters based on user support sets and popularity biases, respectively. Concurrently, contrastive learning has emerged as a powerful tool to align disparate feature spaces; for instance, CCFCRec~\cite{zhou2023contrastive} bridges the gap between content and collaborative views via mutual information maximization, Aligning Distillation~\cite{huang2023aligning} employs ranking-based distillation to transfer knowledge from warm to cold items, and CMCLRec~\cite{xu2024cmclrec} extends these principles to cross-modal sequential recommendation. Addressing topological limitations, graph neural networks have been adapted for isolated cold nodes, with CC-GNN~\cite{xv2023commerce} introducing semantic phrase nodes to enhance e-commerce search connectivity, and CGRC~\cite{kim2024content} utilizing content-based graph reconstruction to predict missing links. Furthermore, cross-domain recommendation has seen breakthroughs like UEDMCF~\cite{liu2024user}, which leverages optimal transport to map user distributions across non-overlapping domains. Most recently, the field is embracing Large Language Models (LLMs), where ColdLLM~\cite{huang2025large} and LLMTreeRec~\cite{zhang2025llmtreerec} demonstrate the efficacy of zero-shot reasoning and tree-structured retrieval in overcoming extreme cold-start scenarios. Aliboost~\cite{shen2025aliboost} first introduced the boosting principles and recommender techniques to ensure continuous growth of new items. However, these methods either only focus on immediate conversion or fail to accurately quantify the long-term value of new items.

\subsection{Generative Recommendation}

Over the past decade, industrial recommendation has relied on Deep Learning Recommendation Models (DLRM) focused on discriminative ranking. 
Recently, Generative AI and LLMs have catalyzed a shift toward Generative Recommendation~\cite{cui2022m6,geng2022recommendation,rajput2023recommender,zhang2025onetrans}.
In terms of Generative Retrieval, TIGER~\cite{rajput2023recommender} pioneered the use of RQ-VAE for semantic ID quantization, enabling autoregressive item prediction in cold-start scenarios, a direction advanced by OneRec~\cite{deng2025onerec} and TBGRecall~\cite{liang2025tbgrecall}, which integrate preference alignment and session-wise generation for industrial-scale, alongside GPT4Rec~\cite{li2023gpt4rec} which leverages search query generation. Regarding LLM-based Paradigms, P5~\cite{geng2022recommendation} established a unified text-to-text framework for multi-task recommendation via personalized prompts, inspiring subsequent works like VIP5~\cite{geng2023vip5} for multimodal tasks, TALLRec~\cite{bao2023tallrec} for efficient instruction tuning, and NoteLLM~\cite{zhang2024notellm} and RecFormer~\cite{li2023text} for content-centric representation generation. To support the computational demands of these generative models, Model Architectures are undergoing significant evolution; notably, Meta's HSTU~\cite{zhai2024actions} introduced a hierarchical sequential transduction unit that replaces standard attention to validate scaling laws in trillion-parameter recommendation models, a high-efficiency direction echoed by KuaiFormer~\cite{liu2024kuaiformer} and MTGR~\cite{han2025mtgr} in real-time industrial applications. Inspired by these works, we adopt a generative approach integrated with multi-value alignment.

\section{Methodology}

\begin{figure*}[t]
  \centering
  \includegraphics[width=\textwidth]{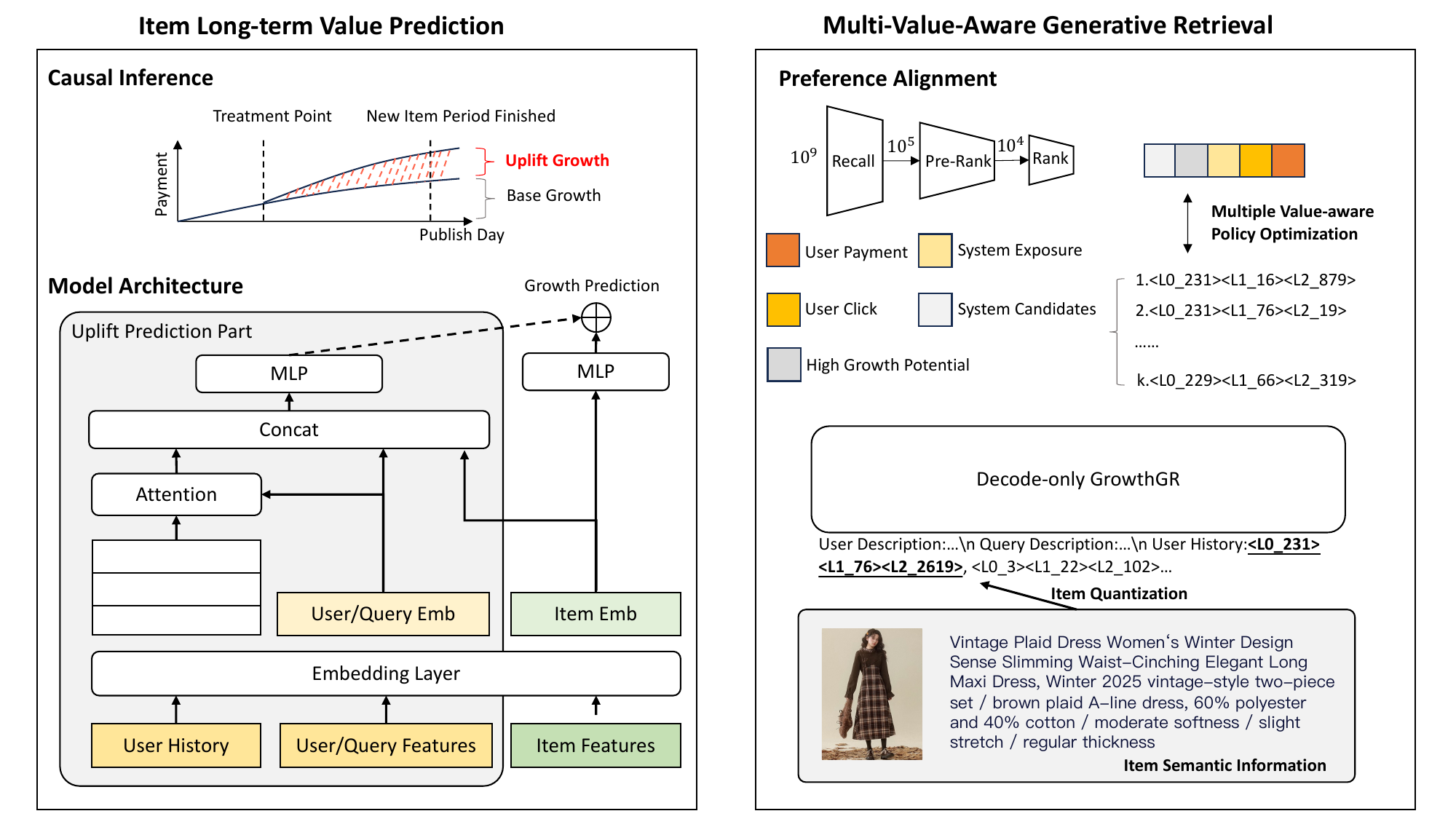}
  \caption{The overview of the proposed GrowthGR framework. 
  }
  \label{overall}
\end{figure*}

In this section, we present the technical details of GrowthGR, a comprehensive framework designed to foster sustainable new item growth through multi-value-aware generative retrieval, as illustrated in Figure~\ref{overall}. The methodology is structured into two primary components: 1) Item Long-term Transaction Value Prediction (ItemLTV): We describe our causal inference approach to quantify the uplift effect of new items. This module estimates the potential transaction increment triggered by user interactions.
2) Multi-Value-Aware Generative Retrieval (MultiGR): We detail the generative architecture and our proposed training strategy to balance immediate conversion value with the predicted
long-term potential.

\subsection{Item Long-term Transaction Value Prediction}

The ItemLTV module identifies the growth potential of new items by modeling the causal impact of user interactions. 
In this context, we posit that user interactions facilitate item conversion from two perspectives. First, at the system level, these interactions enhance the item's distribution efficiency in a personalized manner. For instance, if a new dress is clicked by a college student, the system leverages this interaction to distribute the item to more students with similar user profiles. Second, at the user level, interactions such as conversions directly influence the item's front-end display, which in turn affects the purchasing decisions of users in future exposures. Thus, we evaluate the impact of these two effects by employing a Counterfactual Causal Inference framework to estimate the long-term transaction increment triggered by specific user engagements. 

\subsubsection{Problem Definition}
In our framework, we define the Treatment Point as the occurrence of a user click ($W_i = 1$). Let $X_i = \{x_i^I, x_i^C\}$ denote the covariates, where $x_i^I$ represents item-specific attributes and $x_i^C$ captures the user/query context. We define the potential outcomes for an item $i$ as:
\begin{itemize}
    \item $Y_i(1)$: The average daily order count during the subsequent 7-day window following the initial 30-day "New Item Period" if a click occurs ($W_i = 1$).
    \item $Y_i(0)$: The base order count over the same 7-day period without the specific click ($W_i = 0$).
\end{itemize}
Based on them, we further define the Uplift Growth as $Y_i(1)-Y_i(0)$ and the Base Growth as $Y_i(0)$. We specifically choose user click over system exposure as the treatment variable for three critical reasons: 1) \textbf{Enhancing Signal Reliability}: System exposure alone has a limited effect on an item's inherent growth trajectory and introduces high predictive variance in large-scale environments. 2) \textbf{Mitigating Negative Feedback}: In industrial search engines, exposure without a subsequent click is interpreted as negative feedback, leading the system to suppress the item in future pre-rank and rank stages. 3) \textbf{Reinforcing Success Cycles}: User clicks followed by downstream actions, such as favoriting or purchasing, provide high-fidelity labels. This feedback is returned to the model architecture, enhancing future efficiency estimations and naturally increasing the item's future exposure opportunities.

To account for the heavy-tailed distribution of transaction values, we perform estimation in the logarithmic space. Our objective is to estimate the Conditional Average Treatment Effect (CATE), denoted as $\tau(X_i)$, representing the uplift growth attributable to the click:
\begin{equation}
    \tau(X_i) = E[\log(Y_i(1)+1) - \log(Y_i(0)+1) | X_i].
\end{equation}

\subsubsection{Two-Tower Model Architecture}
 We employ a two-tower structure to estimate these components:
\begin{itemize}
    \item \textbf{Item Tower (Base Growth)}: This branch processes Item Features (e.g., semantic attributes) through an Embedding Layer to generate an Item Embedding. This embedding is passed through a dedicated MLP to output the predicted score for the item's base growth $G_{base}$:
    \begin{equation}
        G_{base}(X_i)=f_1(x_{i}^I).
    \end{equation}
    
    \item \textbf{Uplift Prediction Tower (Uplift Growth)}: This tower integrates User History, User/Query Features, and Item Embedding. It utilizes an Attention mechanism to capture fine-grained interactions between user intent and item attributes, followed by a concatenation operation and an MLP to estimate the incremental value $\tau(x_i)$:
    \begin{equation}
        G_{uplift}(X_i)=f_2(\left[g(x_{i}^C), x_{i}^I\right]).
    \end{equation}
\end{itemize}

During the training phase, the model is optimized based on the observed interaction. 
The predicted log-outcome $\hat{y}_i$ is formulated based on the treatment assignment $W_i$:
\begin{equation}
    \hat{y}_i = G_{base}(X_i) + W_i \cdot G_{uplift}(X_i).
\end{equation}
Then, we 
employ Mean Squared Error (MSE) as the loss function for model optimization:
\begin{equation}
    L_{ItemLTV} = \sum_i||\hat{y}_i-\log (Y_i+1)||^2_2,
\end{equation}
where $Y_i$ is the ground truth order count. By supervising the combined output against actual purchase data, the model learns to decouple the item's base growth value from the value added by user interaction.

\subsection{Multi-Value-Aware Generative Retrieval}

The MultiGR module is designed to identify high-potential new items from a billion-scale candidate pool by shifting the retrieval paradigm from traditional ID-matching to semantic ID generation, with an emphasis on balancing immediate conversion value and long-term value, to overcome data sparsity and ensure sustainable item discovery. 

\subsubsection{Item Quantization}
To represent items in a discrete semantic space, we employ a hierarchical tokenization strategy inspired by the TIGER framework. We first utilize a pre-trained e-commerce foundation model to generate unified multimodal representations of items, capturing fine-grained item semantic information such as titles, properties, and images. We apply a Residual Quantized Variational AutoEncoder (RQ-VAE) \cite{lee2022autoregressive} to these representations to derive a three-layer semantic ID. Each item is assigned a hierarchical path (e.g., $<L0\_231><L1\_16><L2\_879>$), which preserves structural relationships between items via shared hierarchical prefixes and allows the model to generalize across semantically similar items even without historical interaction data. 

\subsubsection{Generative Architecture}
The core of MultiGR is a Decoder-only Transformer-based model that formulates retrieval as a sequence-to-sequence task. The details are:
\begin{itemize}
    \item \textbf{Input Composition}: The model takes a concatenated natural language string as input, including User/Query Features, description queries, and the user's historical interaction sequence represented as semantic IDs.
    \item \textbf{Autoregressive Generation}: The model learns to generate the semantic ID of the next item, effectively modeling the underlying user interests and transition patterns.
    \item \textbf{Constrained Decoding}: During inference, we implement a trie-based constrained decoding to ensure that the generated tokens follow a valid path within the pre-defined SID hierarchy, effectively filtering candidate items within the generation process.
\end{itemize}

\subsubsection{Multi-Objective Training Strategy}
To balance immediate conversion efficiency with long-term ecological value, we propose a two-stage training strategy:

\textbf{Supervised Pre-training.} The model is first trained using the Next Token Prediction (NTP) task on historical transaction logs. The training loss function is as follows:
\begin{equation}
    \mathcal{L}_{NTP} = -\frac{1}{\sum_{k=1}^{N}|o_k|}\sum_{i=1}^{N}\sum_{t=1}^{|o_i|} \log \mathcal{P}(o_{i,t} | o_{i,<t}, x),
\end{equation}
where $N$ is the number of samples, $x$  denotes the input context (e.g., behavioral sequences), and $\mathcal{P}$ denotes the probability of decoding the semantic ID $o_i=(o_{i,1},\dots,o_{i,|o_i|})$ corresponding to the purchased item.
This stage provides a foundational capability for well-formed generation and aims to help the model learn the collaborative filtering signals of the platform while aligning latent representations of semantic IDs with natural language descriptions.

\textbf{Preference Alignment.} In the prior training stage, reliance on random negative sampling and the Maximum Likelihood Estimation (MLE) objective makes it difficult for the model to distinguish fine-grained relative rankings across outcomes in the causal funnel (e.g., exposure → click → purchase). Furthermore, since the inference paradigm of generative retrieval is typically beam search, it is crucial to focus not only on the scoring capability for top candidates but also on the collective value of the retrieved set. We propose MoPO, which inherits the efficient group-relative optimization framework of GRPO \cite{shao2024deepseekmath} while extending it with a Multi-value Reward Engine. Unlike vanilla GRPO that treats rewards as independent feedback, MoPO calibrates the reward signal using a clipped importance weighting (CIW) mechanism to counteract popularity bias and models the cascaded search funnel explicitly.

We consider the following objectives: 
\begin{itemize}
    \item \textbf{Cascaded Value Labels}: We utilize a hierarchy of labels including user purchase, user click, system exposure, and system rank candidates to model the search funnel.
    \item \textbf{Long-term Labels}: Since the ItemLTV module estimates the uplift triggered by user interactions, it is essential to ensure that the items exhibit a high predicted click-through rate. Thus, we leverage the ItemLTV module to identify high growth potential items by selecting those clicked items whose predicted uplift exceeds the global mean.
\end{itemize}
Different scores are assigned to different interaction labels, and the total reward is a calibrated weighted sum of these scores:
\begin{equation}
    r_i = \operatorname{Clip}(-\log\pi_{\theta_{old}}(o_{i}|x),1,M)\sum_{k}w_k s_k,
\end{equation}
where $s_k\in\{0, 1\}$ is the indicator for the k-th objective (e.g., purchase), $w_k$ is its corresponding weight, and $\pi_{\theta_{old}}(o_{i}|x)$ is the probability of generating the semantic ID $o_{i}$ under the behavior policy. The hyperparameter $M>1$ serves as an upper bound. To align the generated probability distribution with the ground-truth value distribution, we utilize the label distribution of the samples to determine the weights $\{w_k\}$. Furthermore, we incorporate a clipped importance weight, defined as $\operatorname{Clip}(-\log\pi_{\theta_{old}}(o_{i}|x),1,M)$, to effectively re-weight the reward signals. By employing the negative log-likelihood, we naturally assign higher importance to rare, long-tail semantic IDs that are assigned low probability by the behavior policy under the current context, while compressing the dynamic range of weights to ensure training stability. The clipping mechanism further ensures that head IDs retain a baseline reward to prevent vanishing gradients, while capping the influence of extreme outliers. This strategy effectively prevents the model from being dominated by head samples, encouraging it to capture rare but high-value patterns and ultimately leading to a more diverse and optimal retrieval set.
Let $\rho_{i,t}(\theta) = \frac{\pi_\theta(o_{i,t}|x,o_{i,<t})}{\pi_{\theta_{old}}(o_{i,t}|x,o_{i,<t})}$ denote the importance sampling ratio. The final loss function is as follows:
\begin{equation}
    \mathcal{L}_{MoPO} = -\frac{1}{G} \sum_{i=1}^G \frac{1}{|o_i|} \sum_{t=1}^{|o_i|}\left[   \mathcal{J}_{i,t}^{CLIP}(\theta)  - \beta D_{KL}(\pi_\theta || \pi_{ref}) \right],
\end{equation}
where we define the clipped surrogate objective as: 
\begin{equation}
    \mathcal{J}_{i,t}^{CLIP}(\theta)= \min\left[\rho_{i,t}(\theta)\hat{A}_{i,t}, \text{clip}(\rho_{i,t}(\theta), 1-\epsilon, 1+\epsilon)\hat{A}_{i,t}\right],
\end{equation}
and $\hat{A}_{i, t}=\frac{r_i-\operatorname{mean}(\mathbf{r})}{\operatorname{std}(\mathbf{r})}$ is the normalized advantage.

\subsubsection{Inference}
At inference time, MultiGR leverages beam search to retrieve the top-\textit{k} sequences with the highest likelihood. Furthermore, a constrained decoding strategy is applied to guarantee that the generated semantic ID paths strictly conform to the pre-defined valid candidates. To resolve semantic ID collisions where multiple items share the same identifier, we utilize a dense retrieval model for internal re-ranking. Please refer to Appendix~\ref{decoding_strategy} for more details.


\section{Experiments}

In this section, we conduct a series of experiments on Taobao's real-world e-commerce data to comprehensively validate the effectiveness, the component contributions, the scaling properties, and the industrial applicability of GrowthGR. 
We aim to address the following research questions:

\begin{itemize}
    \item \textbf{RQ1:} How does GrowthGR perform compared to competitive retrieval baselines?
    \item \textbf{RQ2:} How do the individual components contribute to capturing long-term value and improving retrieval accuracy?
    \item \textbf{RQ3:} Does the proposed generative retrieval framework adhere to scaling laws?
    \item \textbf{RQ4:} How does GrowthGR balance immediate conversion efficiency with the long-term growth of new items on Taobao?
\end{itemize}




\subsection{Experimental Setup} 

\subsubsection{Datasets}
We conduct extensive experiments on Taobao’s real-world e-commerce search dataset, which comprises billions of user-item interactions. Taobao is a leading global platform where intelligent retrieval solutions must handle massive-scale data to serve millions of users and items daily. 

We utilize two distinct datasets tailored for our Multi-Value-Aware framework: 1) Uplift Prediction Dataset: Given that our ItemLTV module focuses on long-term transaction value growth triggered by specific user engagements, we utilize a specialized click-oriented dataset. It contains 2.4 billion interactions collected from historical search logs, covering 0.1 billion users and 0.16 billion items including 3 million new items. It is primarily used to train the uplift prediction model. 2) Full-chain Conversion Dataset: This dataset is designed to train the foundational retrieval and collaborative filtering capabilities within MultiGR. This set contains 5.0 billion multi-behavior interactions spanning 0.17 billion users and 0.11 billion items including 2 million new items. This large-scale data provides a solid foundation for the model to learn the basic semantic and behavioral alignments between users and items. 

All online experiments are conducted on Taobao’s primary online search system. This environment supports large-scale validation of sustainable growth metrics. The daily traffic includes approximately 1.6 billion interactions involving 0.15 billion active users and 0.3 billion unique items, of which 3 million are new items.

\begin{table}
  \caption{Statistics of the Datasets used in GrowthGR.}
  \label{datasets}
  \begin{tabular}{lcccc}
    \toprule
    \textbf{Dataset} & \textbf{Interactions} & \textbf{Users} & \textbf{Items} & \textbf{New Items} \\
    \midrule
    Uplift & 2.4B & 0.1B & 0.16B & 3M \\
    Full-chain & 5.0B & 0.17B & 0.11B & 2M \\
    Online Daily & 1.6B & 0.15B & 0.3B & 3M \\
    \bottomrule
  \end{tabular}
\end{table}

\subsubsection{Evaluation Metrics} 
To verify the effectiveness of GrowthGR in balancing immediate conversion efficiency with sustainable new item growth, we utilize a comprehensive set of offline and online metrics. 

Offline Metrics. Our evaluation considers multiple ground-truth labels, including In-scenario Conversions marked as \textit{search} (to assess the retrieval of head, mid, and tail items within the search context), Platform-wide Conversions marked as \textit{all-net} (to measure the model’s ranking capability beyond the specific search scene, capturing incremental cross-scenario relevance) and Long-term Value marked as \textit{long-term} (to assess the fidelity of the retrieval results to the ItemLTV module’s uplift predictions). Specifically, all-net labels are derived by linking a user’s site-wide transactions based on query relevance. Empirically, the performance on search labels reflects the model's ability to fit the distribution of the current system, serving as an indicator of its alignment with the existing production pipeline. Recall performance on all-net labels exhibits higher consistency with final online immediate outcomes, thereby providing a more unbiased reflection of the model's search effectiveness, while long-term labels follow the same definition as in Preference Alignment where they are empirically shown to be positively correlated with the future conversion of new items online. We employ two standard ranking metrics to evaluate the retrieval performance on historical logs:
\begin{itemize}
    \item \textbf{Recall@$k$}: This metric measures the model's ability to recall relevant items within the top $k$ candidates. 
    The formula is defined as:
    \begin{equation}
        \text{Recall}@k = \frac{1}{N} \sum_{i=1}^{N} \mathbb{I}(\text{rank}_i \le k),
    \end{equation}
    where $N$ is the total number of relevant items, $\text{rank}_i$ is the predicted rank and $\mathbb{I}(\cdot)$ is the indicator function.
    \item \textbf{Normalized Discounted Cumulative Gain (NDCG)} \cite{jarvelin2002cumulated}: 
    To evaluate the quality of the ranked list and the model's alignment with user preferences, we use NDCG. 
\end{itemize}

Online Metrics. The online performance is measured via online A/B testing on Taobao, focusing on transaction impact, traffic structure, and long-term growth:
\begin{itemize}
    \item \textbf{Gross Merchandise Volume (GMV)}: The total sales revenue generated within the search scenario. This is the core metric for measuring immediate conversion efficiency and the platform's overall economic health.
    \item \textbf{Page View Ratio (PVR)}: Defined as the proportion of total exposures allocated to new items. This metric reflects the traffic structure and the system's success in distributing traffic to new items to mitigate the ``Matthew effect''.
    \item \textbf{Transaction Impact at T+30 (TI@30)}: We adopt daily GMV during the subsequent 7-day window following the initial 30-day period as a proxy for the successful transition of items from the exploration phase to the exploitation phase. A higher TI@30 indicates that our model can accurately identify potential items during their early exposure window and provide them with an optimized growth trajectory. This aims to ensure that the platform does not merely provide temporary traffic boosts but fosters genuine market fit and sustained transaction volume for new items.
\end{itemize}

\subsubsection{Implementation Settings}

The Dense Retrieval (DR) model employed in our experiments consists of 40B sparse embedding parameters and 10M dense network parameters, which also serves as the current state-of-the-art production model. All generative models are configured with a total size of 0.5B, including 0.15B embedding parameters.


\subsection{Main Results (RQ1)}

\subsubsection{ItemLTV Effectiveness}

\begin{table}[t]
    \centering
    \caption{Performance Comparison of ItemLTV.}
    \label{tab:uplift}
    \begin{tabular}{lcc}
        \toprule
        Metric & MSE ($\downarrow$) & NDCG ($\uparrow$) \\
        \midrule
        base  & 1.348 & 0.842 \\
        +uplift part & \textbf{1.329} & \textbf{0.853} \\
        \bottomrule
    \end{tabular}
\end{table}

\begin{table*}[ht]
    \centering
    \caption{Experimental performance comparison on the Taobao real-world dataset. Bold indicates the best results while underline denotes the second-best results.}
    \label{tab:main_results}
    \resizebox{\textwidth}{!}{
    \begin{tabular}{lcccccccccccc}
        \toprule
        & \multicolumn{4}{c}{\textbf{search}} & \multicolumn{4}{c}{\textbf{all-net}} & \multicolumn{4}{c}{\textbf{long-term}} \\
        \cmidrule(lr){2-5} \cmidrule(lr){6-9} \cmidrule(lr){10-13}
        Method & Recall@10 & Recall@100 & Recall@1000 & NDCG & Recall@10 & Recall@100 & Recall@1000 & NDCG & Recall@10 & Recall@100 & Recall@1000 & NDCG \\
        \midrule
        DR               & \underline{0.4115}      & \underline{0.6775}      & \underline{0.8820}      & \underline{0.3628}      & 0.2463      & 0.4260      & 0.6194      & 0.2422      & 0.3990      & 0.6704      & 0.8772      & \underline{0.4119}      \\
        TIGER             & 0.2950 & 0.5909 & 0.8215 & 0.2842 & 0.2159 & 0.4430  & 0.6969  & 0.2307 & 0.3810 & 0.7209 & 0.8883 & 0.3375 \\
        GrowthGR          & 0.3419 & 0.6306 & 0.8438 & 0.3160 & \underline{0.2568} & \underline{0.4981}  & \underline{0.7147}  & \underline{0.2555}  & \underline{0.4440} & \underline{0.7523} & \underline{0.8991} & 0.3820 \\
        GrowthGR-twoStage & \textbf{0.513}  & \textbf{0.8062}  & \textbf{0.8820} & \textbf{0.4102}  & \textbf{0.3187}  & \textbf{0.5702}  & \textbf{0.7578}  & \textbf{0.2970}  & \textbf{0.6221}  & \textbf{0.8558}  & \textbf{0.9261}  & \textbf{0.4801}  \\
        \bottomrule
    \end{tabular}
    }
\end{table*}

We first conduct an offline evaluation of the ItemLTV module independently. Since the uplift effect of user interactions on items is inherently counterfactual and lacks direct ground truth for evaluation, we assess the module's performance by its ability to fit the aforementioned average daily order count of new items using MSE for regression accuracy and NDCG for ranking performance. Specifically, the "Base" model relies solely on the item-side tower for label prediction, while the "+uplift part" configuration incorporates the uplift prediction component for joint estimation. As shown in Table~\ref{tab:uplift}, the results demonstrate that the overall prediction reaches a high performance level. Notably, the "+uplift" variant exhibits superior fitting and ranking capabilities, confirming that the uplift module effectively captures the impact of user interactions. 

Further analysis demonstrates that new items are significantly more sensitive to user interactions in their early stages (see Appendix~\ref{listing} for a comprehensive discussion).


\subsubsection{MultiGR Effectiveness}

Table~\ref{tab:main_results} summarizes the offline performance of our proposed models compared to state-of-the-art baselines across three distinct labels: search, all-net, and long-term. Specifically, DR is the state-of-the-art dense retrieval model currently deployed in Taobao Search's production system. TIGER \cite{rajput2023recommender} serves as a generative retrieval baseline, implemented using semantic IDs and Supervised Fine-Tuning. GrowthGR-twoStage denotes a two-stage retrieval pipeline where GrowthGR serves as the initial retriever to produce top-2,000 candidates, followed by a re-ranking stage using the DR model to generate the final recommendation list. We observe several key findings as follows.

GrowthGR demonstrates a significant competitive advantage over traditional and generative baselines. GrowthGR consistently outperforms TIGER across all label types, particularly in all-net and long-term metrics. For instance, in the all-net category, GrowthGR achieves a Recall@1000 of 0.7147 compared to TIGER's 0.6969. This highlights that our multi-value-aware approach and preference alignment effectively address the limitations of standard supervised fine-tuning in capturing cross-scenario relevance and growth potential.
While DR
shows strong performance on search labels reflecting its heavy optimization for the existing system's distribution, GrowthGR achieves the second-best results in all-net and long-term recall. Notably, on long-term Recall@1000, GrowthGR (0.8991) outperforms DR (0.8772). This underscores GrowthGR's superior ability to identify items with high growth potential that the current production system might overlook.

The GrowthGR-twoStage variant, which utilizes GrowthGR as the primary retriever followed by DR for re-ranking, achieves the best overall performance across every single metric. These results indicate that GrowthGR provides a high-quality candidate set rich in both immediate relevance and long-term potential. By serving as a robust initial retriever, it allows the ranking stage to operate on a pool of candidates that is fundamentally better aligned with both user intent and platform sustainability.

The performance on long-term labels is particularly telling for the MultiGR module's effectiveness. GrowthGR-twoStage achieves an NDCG of 0.4801 and a Recall@1000 of 0.9261. This exceptionally high fidelity to uplift-based labels demonstrates that our framework successfully translates Preference Alignment into actual retrieval behavior, ensuring that items predicted to have the highest marginal return on exposure are prioritized. Empirically, the strong performance on these labels is positively correlated with the future conversion of new items online, validating our strategy for sustainable ecosystem growth.

\subsection{Ablation Study (RQ2)}

\begin{table}[t]
  \centering
  \caption{Ablation Study of GrowthGR Components.}
  \label{tab:ablation}
  \resizebox{0.47\textwidth}{!}{
  \begin{tabular}{lcccc}
    \toprule
    & \multicolumn{2}{c}{\textbf{all-net}} & \multicolumn{2}{c}{\textbf{long-term}} \\
    \cmidrule(lr){2-3} \cmidrule(lr){4-5}
    Model & Recall@10 & Recall@1000 & Recall@10 & Recall@1000 \\
    \midrule
    \textbf{GrowthGR (Ours)} &  &  &  &  \\
    \quad w/o ItemLTV         & -0.0006          & 0.0004          & -0.0532          & -0.0165          \\
    \quad w/o CIW            & -0.0010          & -0.0038          & -0.0004          & -0.0060          \\
    \quad w/o MoPO            & -0.0220          & -0.0368          & -0.0539          & -0.0325          \\
    \bottomrule
  \end{tabular}
  }
\end{table}

To further investigate the contribution of each component in GrowthGR, we conduct an ablation study. As shown in Table~\ref{tab:ablation}, removing either the ItemLTV module, the CIW or the MoPO strategy leads to a noticeable performance decline across all metrics. 

\subsubsection{w/o ItemLTV}
We conducted an ablation study by removing the scores generated by the ItemLTV module from the MoPO framework. This allows us to investigate its specific impact on the recall performance for both immediate conversion and new items with high long-term value uplift.
Specifically, the absence of ItemLTV results in a drop of 1.65 pt in Recall@1000 and 5.32 pt in Recall@10 for long-term labels, validating its critical role in capturing long-term value. Meanwhile, we observed that the impact on the Recall@1000 for immediate conversion was negligible, with a marginal decrease of only 0.04 pt.

\subsubsection{w/o CIW}
We further investigated the impact of the clipped importance weighting mechanism by excluding the weighting term. The absence of CIW leads to a consistent performance degradation across all metrics. Specifically, we observed a decrease of 0.38 pt in Recall@1000 for all-net labels and a more pronounced drop of 0.60 pt in Recall@1000 for long-term labels. These results confirm that CIW is instrumental in mitigating popularity bias. By calibrating the rewards of frequently sampled head items, CIW encourages the model to explore and prioritize high-value yet sparse patterns in the long tail. 

\subsubsection{w/o MoPO}
Similarly, we replaced the MoPO module with vanilla GRPO targeting immediate conversion as a baseline to evaluate the superiority of our proposed training paradigm. The results demonstrate that MoPO significantly boosts Recall@10 by 2.2 pt and Recall@1000 by 3.68 pt for all-net labels. Meanwhile, an increase of 5.39 pt in Recall@10 and 3.25 pt in Recall@1000 is observed for long-term labels. The performance degradation observed in the variant without MoPO confirms that MoPO is essential for achieving optimal retrieval accuracy.

To examine the impact of aligning training-time sampling with inference-stage decoding, we further explore different rollout strategies in Appendix~\ref{rollout}.

\subsection{Scaling Analysis (RQ3)}

\begin{figure}[t]
  \centering
  \includegraphics[width=\linewidth]{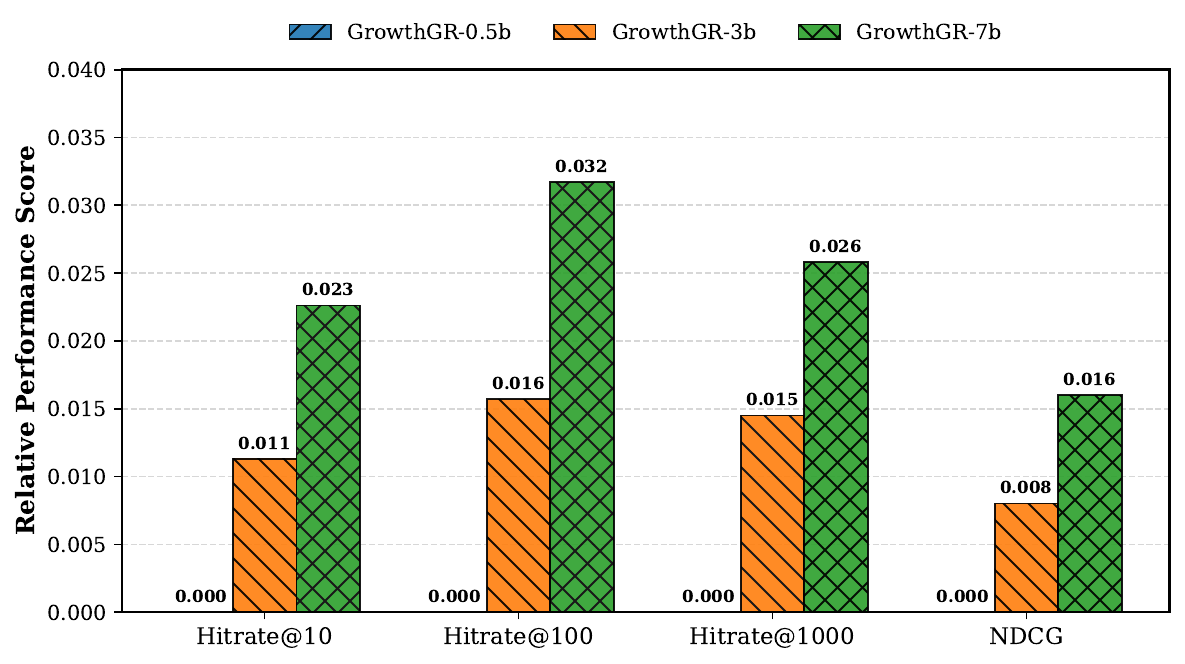}
  \caption{Performance improvement on all-net labels across model scales relative to the 0.5B baseline.}
  \label{fig:scale}
\end{figure}

To evaluate the scalability of the proposed GrowthGR framework, we conduct a systematic investigation by varying the model capacity across three parameter scales: 0.5B, 3B, and 7B. This analysis aims to determine whether our architectural design can effectively leverage increased computational power and parameters.

As illustrated in Figure~\ref{fig:scale}, GrowthGR exhibits a clear and consistent scaling trend. We observe a monotonic improvement across all evaluation metrics, including $\text{Recall@}k$ and NDCG as the parameter count increases. Notably, shifting from 0.5B to 7B yields substantial relative gains (e.g., over +2.0 pt in Recall@1000), suggesting that the model has not yet reached a performance plateau. These empirical results underscore that GrowthGR adheres to the scaling law, demonstrating its potential for further performance breakthroughs as model capacity continues to expand in large-scale industrial applications.

\subsection{Online A/B Testing (RQ4)}

We deployed GrowthGR on Taobao, a leading e-commerce platform, with billion-scale daily traffic for an observation period exceeding two months. In the Taobao search system, new items retrieved by the GrowthGR framework undergo a lightweight re-ranking process and are then output as a dedicated new item channel. This stream is subsequently blended with other streams (e.g., advertisement, mainstream) before being presented to users. Specific implementation details of the online deployment are available in Appendix~\ref{deploy}.

Our experimental design considers both immediate conversion efficiency and the long-term ecological impact on new items through two distinct A/B testing methodologies: 
\begin{itemize}
    \item \textbf{User-side A/B Testing}: We employ a standard user-level partitioning approach where incoming traffic is hashed by user IDs into orthogonal buckets. This setup allows us to compare the immediate conversion efficiency of different retrieval strategies by observing real-time user-item interactions.
    \item \textbf{Item-side A/B Testing}: Recognizing that the growth of new items is a long-term process, we implement item-level partitioning based on a uniform hash of item IDs. Different strategies are applied to specific item buckets to observe the differential growth trajectories over an extended period. To ensure that a specific group of new items is influenced by only one strategy and to prevent cross-contamination between policies, we utilize a coupled bucket mechanism. Each item bucket is exclusively mapped to a corresponding user bucket.
\end{itemize}


\subsubsection{User-side A/B Testing}

To evaluate the real-world impact of GrowthGR, we conducted a large-scale online A/B experiment. The results demonstrate that GrowthGR significantly enhances both the exposure and conversion efficiency of new items, yielding a 5.39\% increase in new item GMV and a 1.54\% lift in PVR. From an ecosystem perspective, GrowthGR effectively orchestrates the dynamics of our two-sided marketplace: it satisfies the evolving novelty-seeking behavior of users while simultaneously providing tangible incentives for merchants to launch innovative items. Notably, the overall Search GMV grew by 0.31\%, a statistically significant result. This confirms that our strategy does not lead to traffic cannibalization or a zero-sum redistribution. Instead, it generates genuine incremental value, thereby bolstering the platform's aggregate revenue.

We also conducted a further analysis of the impact across different categories, with detailed results provided in Appendix~\ref{category}

\subsubsection{Item-side A/B Testing}

To further investigate the long-term impact on the lifecycle of new items, we conducted an item-side A/B test. Specifically, we tracked the performance of new items and compared their transaction scales after the 30th day post-launch period between the treatment group and the control group. 

The experimental results reveal that new items supported by the GrowthGR strategy achieved a +20.0\% uplift in TI@30. This significant improvement demonstrates that our framework does not merely provide a transient "injection" of traffic during the initial cold-start phase; instead, it fundamentally enhances the long-term traffic-carrying capacity and conversion efficiency of new items within the search ecosystem. From a strategic perspective, by successfully nurturing high-potential items into mature, high-performing ones, GrowthGR ensures a continuous supply of fresh content that aligns with evolving user preferences. This fosters a more dynamic and sustainable merchant ecosystem on Taobao, effectively bridging the gap between niche user needs and emerging item trends.

\section{Conclusion}

In this paper, we addressed the critical challenge of fostering sustainable growth for new items in large-scale e-commerce search scenarios. Industrial search systems are often plagued by the ``Matthew Effect'', where traditional retrieval models favor head items with abundant historical data, leaving new items characterized by massive scale and extreme data sparsity struggling for visibility. To overcome these limitations, we proposed GrowthGR, a novel Multi-Value-Aware Generative Retrieval framework which can accurately quantify the incremental transaction value an item gains relative to its base growth and harmonize immediate conversion efficiency with predicted long-term potential. Extensive offline experiments and online A/B tests on Taobao involving billions of interactions demonstrate that GrowthGR significantly improves the discovery and growth rate of new items without sacrificing overall platform efficiency. Our experimental design, which integrates both user-side and item-side A/B testing, confirms the framework's capability to mitigate popularity bias and support long-term ecosystem health.

\bibliographystyle{ACM-Reference-Format}
\bibliography{sample-base}

\appendix

\section{System Implementation and Deployment}

\label{deploy}

\begin{figure}[t]
  \centering
  \includegraphics[width=\linewidth]{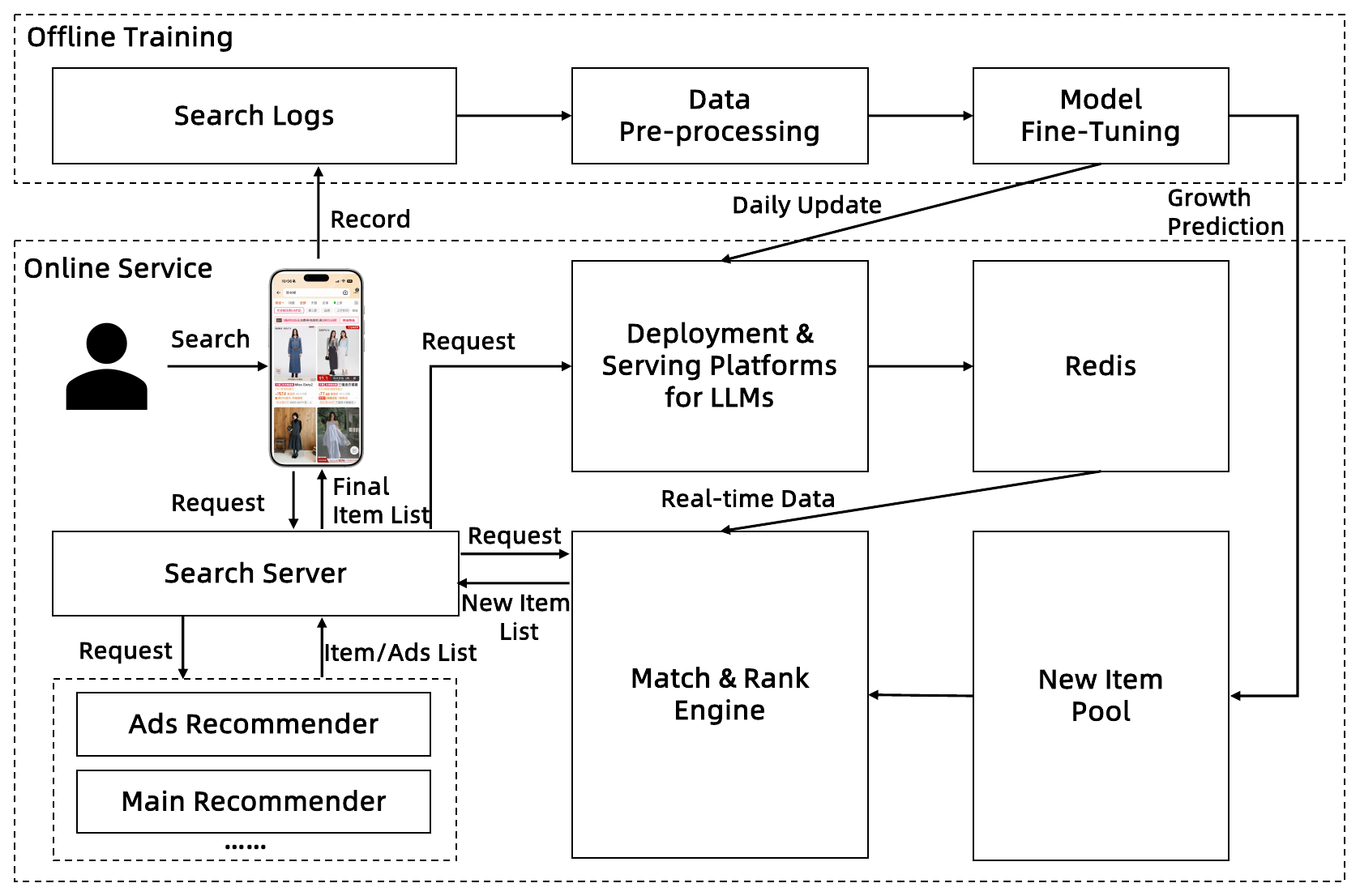}
  \caption{The deployment architecture of GrowthGR.}
  \label{fig:deployment}
\end{figure}

In this section, we detail the system implementation and deployment architecture of GrowthGR, as illustrated in Figure~\ref{fig:deployment}. The system is structured into two primary components: the offline training pipeline and the online service engine.

\subsection{Offline Training Pipeline}
The offline component focuses on high-throughput data processing and continuous model evolution. 

\subsubsection{Data Management}
User search logs are captured via real-time tracking and stored in Alibaba's MaxCompute Platform. These raw logs are pre-processed into structured datasets tailored for the two-tower and generative architectures.

\subsubsection{ItemLTV Training}
To capture long-term transaction dynamics, the ItemLTV module utilizes a daily retrospective labeling mechanism. For each training cycle, transaction labels from the most recent day are paired with historical features from the preceding month to optimize the uplift model. 

\subsubsection{MultiGR Fine-Tuning}
The MultiGR model undergoes daily fine-tuning. It integrates the inference results from ItemLTV with cascaded behavioral labels to ensure the model’s distribution remains aligned with shifting online user preferences. To balance retrieval performance with serving efficiency, we employ a 0.5B parameter configuration for the transformer-based decoder. Once trained, the MultiGR model is deployed to Alibaba's specialized LLM platform for real-time serving.

\subsubsection{New Item Pool Selection}
To optimize traffic allocation, we utilize the item tower from the ItemLTV module to generate growth potential scores for candidate items. From a vast influx of over 200 million new items, the system curates a new item pool of approximately 2 million high-potential items daily.

\subsection{Online Service Engine}
The online engine is designed for low-latency, high-concurrency retrieval and ranking. Upon a user search request, the search server concurrently triggers multiple recommender streams, including advertisement, mainstream, and the specialized new item stream. For the new item stream, the MultiGR model performs inference in parallel with other pre-processing tasks (such as query parsing) and asynchronously writes the retrieved candidates into a Redis cache. When the request reaches the Match \& Rank Engine, it fetches the generative results from Redis. Due to the early-trigger mechanism and the efficiency of the 0.5B model, we achieve an average cache hit rate of 99\% at this stage. A lightweight ranking model then filters the 1,000 candidates to produce a new item list containing the top 20 items. These 20 items are returned to the search server, where they are blended with other results to form the final item list presented to the user.

\section{Decoding Strategy}

\label{decoding_strategy}

\begin{figure}[t]
  \centering
  \includegraphics[width=\linewidth]{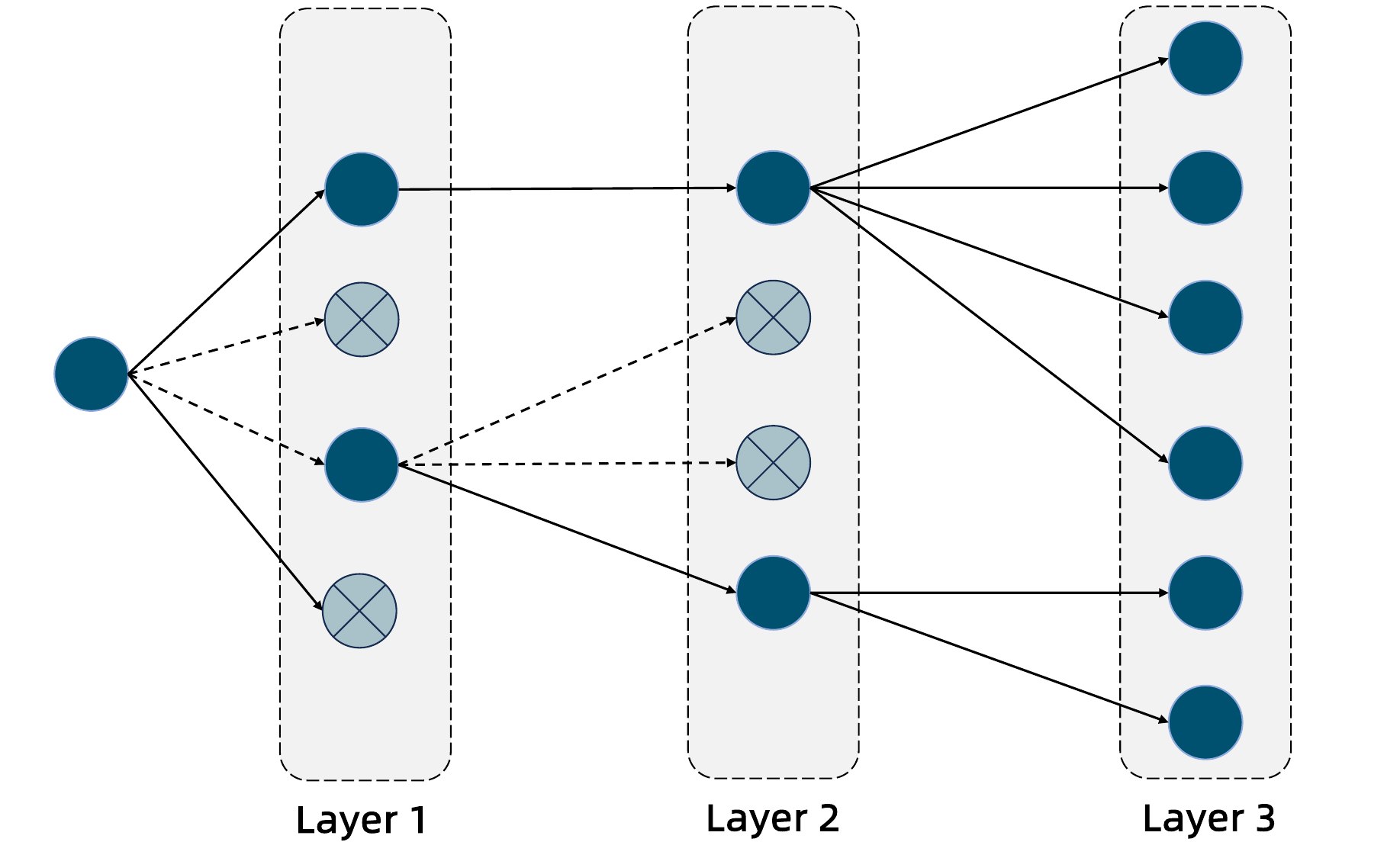}
  \caption{The framework of the decoding strategy.}
  \label{fig:decoding}
\end{figure}

As illustrated in Figure~\ref{fig:decoding}, the decoding strategy primarily incorporates two techniques:

\begin{itemize}
    \item \textbf{Constrained Decoding}: Since the total number of paths in the codebook exceeds the number of valid items, we apply a masking mechanism during the decoding of each token. This mask filters out semantic ID paths that do not correspond to any physical items, effectively excluding them from the logit ranking process.
    \item \textbf{Beam Search}: For offline inference, we employ a standard beam search to identify the top-$k$ semantic IDs with the highest probabilities under the constraints. For online inference, to balance computational efficiency with retrieval performance, we implement a dynamic-width beam search. This approach involves reducing the beam width in the initial layers of the hierarchy to meet latency requirements while ensuring a sufficient number of candidate items are retrieved.
\end{itemize}

\section{More Analysis}

\subsection{Impact of Online Days on Uplift}

\label{listing}

\begin{figure}[t]
  \centering
  \includegraphics[width=\linewidth]{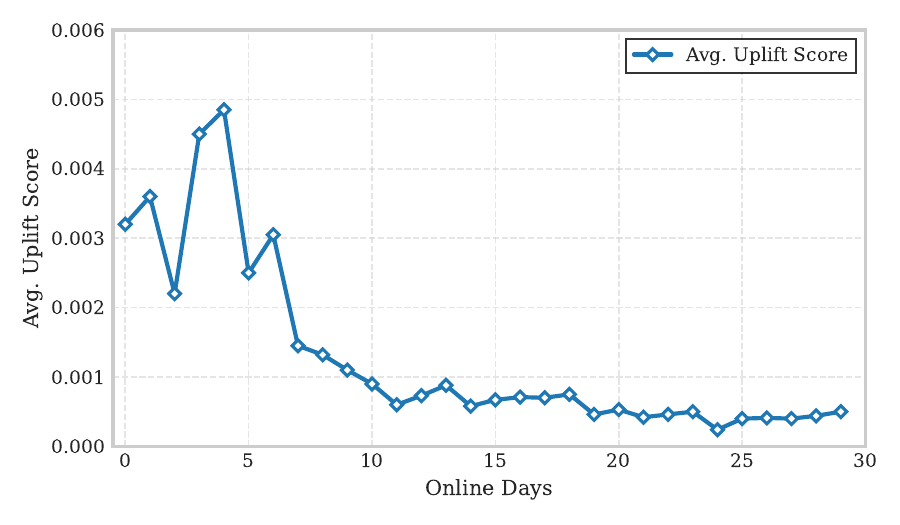}
  \caption{Average Uplift Score Across Different Online Days.}
  \label{fig:uplift_day}
\end{figure}

We analyze the distribution of uplift scores relative to the days since listing. As shown in Figure~\ref{fig:uplift_day}, the findings reveal that the uplift effect per interaction diminishes as the listing time increases. This indicates that new items are more sensitive to user interactions in their early stages, which aligns with empirical intuition in the e-commerce domain.

\subsection{Analysis of Rollout Strategies}
\label{rollout}

\begin{table}[t]
  \centering
  \caption{Performance of different rollout strategies.}
  \label{tab:rollout}
  \resizebox{0.47\textwidth}{!}{
  \begin{tabular}{lcccc}
    \toprule
    & \multicolumn{2}{c}{\textbf{all-net}} & \multicolumn{2}{c}{\textbf{long-term}} \\
    \cmidrule(lr){2-3} \cmidrule(lr){4-5}
    Strategy & Recall@10 & Recall@1000 & Recall@10 & Recall@1000 \\
    \midrule
    Top-p Sampling  & 0.2568 & \textbf{0.7147} & 0.4440 & 0.8991 \\
    Beam Search         & \textbf{0.2722} & 0.7138          & \textbf{0.4459}          & \textbf{0.9064}          \\
    \bottomrule
  \end{tabular}
  }
\end{table}

In this section, we investigate the impact of different rollout strategies during the Preference Alignment of GrowthGR. Since the generative retrieval process relies on a robust exploration-exploitation balance to discover new items with high growth potential, the choice of decoding strategy during training is critical. We specifically compare two widely adopted strategies: Top-p Sampling and Beam Search. For Top-p Sampling, we set $\text{top-}p = 0.95$, $temperature = 1.0$, and $n = 5$. For Beam Search, the beam size $n$ is set to 5. The performance comparison across all-net and long-term metrics is summarized in Table~\ref{tab:rollout}.

Our experimental results reveal a clear trade-off between top-tier retrieval performance and long-tail coverage. In the all-net scenario, Beam Search demonstrates superior performance in top-tier metrics, achieving a Recall@10 of 0.2722, outperforming Top-p Sampling (0.2568). This suggests that Beam Search is highly effective at identifying the most probable candidates, which typically align with head items already favored by the current system distribution. Conversely, Beam Search exhibits comparable yet slightly lower performance in broader retrieval metrics, with Recall@1000 dropping to 0.7138, compared to Top-p Sampling's 0.7147. This decline indicates that Beam Search may marginally suffer from likelihood bias, often neglecting the long-tail candidates that characterize new item distributions.

Interestingly, Beam Search maintains a slight lead in most long-term metrics, such as a Recall@1000 of 0.9064. This is likely because the ItemLTV module provides a strong prior for high-potential items, allowing even exploitative decoding to capture these growth signals effectively.

\subsection{Category-wise Analysis}

\label{category}

\begin{table}[t]
  \caption{Performance comparison across different categories.}
  \label{tab:category_results}
  \centering
  \begin{tabular}{lrr}
    \toprule
    Categories & PV (\%) & GMV (\%) \\
    \midrule
    Apparel \& Fashion & 1.49 & 2.72 \\
    Sports \& Outdoors & 1.45 & 11.98 \\
    Home \& Living & 2.09 & 3.83 \\
    Fast Moving Consumer Goods & 1.47 & 4.05 \\
    3C \& Digital & 1.12 & 22.18 \\
    Stationery, Education \& Fragrance & 2.90 & 0.22 \\
    Health & 1.21 & 0.46 \\
    Food \& Fresh & 2.20 & -0.67 \\
    Jewelry \& Accessories & 1.50 & 8.21 \\
    Automobiles & 1.36 & -0.81 \\
    Pets & 1.20 & -2.62 \\
    Toys \& Hobbies & 0.50 & -5.48 \\
    Home Furnishing & 1.38 & 5.54 \\
    Flowers \& Gardening & 0.65 & 8.11 \\
    Major Appliances & 1.79 & 1.72 \\
    Small Appliances & 1.34 & 21.15 \\
    \bottomrule
  \end{tabular}
\end{table}

The category-wise increments for new items are summarized in Table~\ref{tab:category_results}. These results provide a granular view of how GrowthGR influences the platform's ecosystem. The experimental data reveals several key insights into the behavior of the Multi-Value-Aware retrieval framework:
\begin{itemize}
    \item GrowthGR achieved positive PV increments across all 16 major categories. This demonstrates that the MultiGR module, powered by hierarchical semantic identifiers and generative modeling, effectively bypasses the historical engagement barriers that usually trap new items. Even in categories like Food \& Fresh ($+2.20\%$) and Stationery, Education \& Fragrance ($+2.90\%$), the system successfully injected significant new item traffic into the search results.
    
    \item In several high-value categories, the growth in GMV significantly outpaced the growth in PV. For instance, 3C \& Digital saw a +22.18\% GMV increase with only a $+1.12\%$ PV increase, while Small Appliances achieved $+21.15\%$ GMV growth. 
    By aligning retrieval with these value-rich candidates, the framework does not just blindly boost new items but strategically surfaces those that provide the highest marginal return for the platform.
    
    \item In high-frequency, low-deliberation categories such as Food \& Fresh and Toys \& Hobbies, we observed a slight decline in immediate GMV ($-0.67\%$ and $-5.48\%$ respectively) despite notable PV increases. This phenomenon reflects the inherent challenge of new item distribution: in categories where consumers have extremely strong brand or ``head-item'' loyalty, introducing new candidates can lead to a temporary dilution of immediate conversion efficiency. However, from a sustainable growth perspective, this discovery cost is a necessary investment to prevent category stagnation and maintain long-term seller diversity.
\end{itemize}
By providing a PV increment ranging from $+0.50\%$ to $+2.90\%$, the framework ensures that new items are no longer invisible in the shadow of established head items. The substantial GMV gains in over $70\%$ of the categories further prove that a significant portion of new items are high-quality candidates that were previously under-exposed due to data sparsity. By balancing immediate efficiency with multiple value dimensions, GrowthGR creates a more dynamic and inclusive search environment on Taobao.










\end{document}